\documentclass[11pt,twoside]{article}

\usepackage{asp2006}
\usepackage{fancyhdr,graphicx,caption,rotating,multirow,lscape}
\DeclareGraphicsExtensions{.eps,.eps.gz,.ps,.jpg,.tiff}

\begin{document}

\title{On Using the Rossiter Effect to Detect Terrestrial Planets}

\author{W. F. Welsh and J. A. Orosz}
\affil{Department of Astronomy,
San Diego State University, San Diego, CA, 92182-1221, USA}

\begin{abstract} 
We explore the possibility that the transit signature of an Earth-size 
planet can be detected in spectroscopic velocity shifts via the Rossiter 
effect. Under optimistic but not unrealistic conditions, it should be 
possible to detect a large terrestrial-size planet. While not suitable for 
discovering planets, this method can be used to confirm suspected planets.
\end{abstract}

The Rossiter effect is seen as a radial velocity ``anomaly'' during transit
\citep[see][and the contribution by Winn in these proceedings]{wfwGaudi2007}.
The distortion in the radial velocity curve can be large because the 
effect scales with the rotational velocity of the star which is very much 
larger than its orbital velocity, e.g.\ km/s vs.\ m/s. So even though a small 
fractional area of the star is occulted, the loss of light breaks the 
exquisite symmetry of the photosphere and a distortion of the line profile
occurs. The skewed line profile then gives rise to an apparent radial 
velocity shift. It is important to realize that there is no real change in 
stellar velocity  --- there is only a change in line shape that mimics a change 
in velocity when the line is used to measure the Doppler reflex motion of the 
star. In the case of a terrestrial-like body orbiting at $>$0.5 AU, the 
amplitude of the Rossiter effect can be larger than the radial velocity 
of the host star's orbit. Under such conditions, it is legitimate to ask if 
the Rossiter effect could be used to detect an Earth-like planet.

To investigate the detectability of a transit of a terrestrial-like planet in 
a 1~AU orbit around a Sun-like star, we simulated the Rossiter effect signal 
and then added observational noise. We chose a hypothetical 
telescope+spectrograph configuration that is based on the performance of the 
{\it HARPS} spectrograph  \citep{wfwQueloz2001,wfwMayor2003}
but with much more light--gathering ability. 
Specifically, for a set of short exposures of a V=9 star binned to 60 min, 
we add Gaussian-distributed white noise with $\sigma=0.25$ m/s. This is 
certainly optimistic, but not absurdly so: {\it HARPS} on the 3.6-m telescope 
at La Silla Observatory has achieved $<$0.7 m/s precision in 15 min 
integrations on the 6th magnitude star HD~69830 
\citep{wfwLovis2006}. 
Scaling to 60 min gives $<$0.5 m/s, while scaling to a 10-m telescope gives 
0.25~m/s, assuming photon noise only. Given ample photons, {\it HARPS} can 
reach $\sim$0.2 m/s (Queloz, private communication), as demonstrated when 6--9 
spectra of HD~69830 were averaged together \citep{wfwLovis2006}. 
So the observational noise used here is certainly achievable in principle.
We emphasize that the point of this investigation is to examine the 
feasibility issue: Could a {\it HARPS}--like spectrograph on a 10-m telescope 
(or even better a 30-m telescope!) be able to spectroscopically detect an 
Earth-analog planet? The amplitude of the radial velocity orbit of the host 
star of such a planet would only be $\pm$0.09 m/s spread over 1 year and
therefore much more difficult and time consuming to detect than the
$\sim$0.10--0.30 m/s Rossiter effect signal that occurs over $\sim$13 hours.

It is important to realize that the limiting noise factor is not spectrograph 
or Poisson noise, but intrinsic stellar variability. The stellar variability 
comes in two main sources: periodic stellar pulsations and random red-noise 
``burbling''. The stellar pulsations arise from a superposition of many 
p--mode oscillations in the photosphere, each with an amplitude of $\sim$0.1 
m/s and period of a few minutes. The 5-min solar oscillations are of this 
nature. We model these in the following way, akin to the p--modes seen in 
the sun and other stars, and specifically, like the K1 V star $\alpha$~Cen~B 
\citep{wfwKjeldsen2005}:
We use a total of 60 sinusoids with 15 radial modes (n=17--32) and 
$\ell=0,1,2,3$ angular modes; the amplitude is 0.2 m/s at $\nu_{peak}$ 
and decreases away from the peak as $exp[ - (\nu / \nu_{peak} )^{2} ]$; each 
sinusoid has a lifetime of $3 (\nu/\nu_{peak})$ days over which the phase 
drifts in a random walk manner by an average of $2\pi$. 
The aperiodic stellar variability is produced by convection, granulation, 
microflares, spots, etc., and has a random walk--like characteristic: 
$1/f$ to $1/f^{2}$ power spectrum \citep[e.g.~see][]{wfwPalle1992}.
To simulate the aperiodic red noise, we use a 1st order 
auto-regressive process: $v(t_{i}) = \alpha v(t_{i-1}) + \epsilon_{i}$
where the velocity at time $t_{i}$ equals the previous velocity scaled by 
$\alpha$ and to which a random perturbation $\epsilon$ is given. 
A pure random walk has $\alpha=1$ and power $P(f) \propto f^{-2}$, 
while if $\alpha=0$ the power is white. To insure the velocities do not 
diverge, we use $\alpha=0.995$, resulting in a red power spectrum with 
$P(f) \sim f^{-1.5}$ to $f^{-2}$.
To keep the simulations consistent, the total rms scatter 
caused by the stellar oscillations and burbling is normalized to 0.5~m/s. 
A realization is shown in Fig~1. The darker curve shows the stellar burbling 
while the lighter curve shows the burbling red noise plus stellar 
oscillations.

\begin{figure}[h]
\centering
\begin{minipage}[c]{.49\textwidth}
\includegraphics[width=5cm,angle=-90]{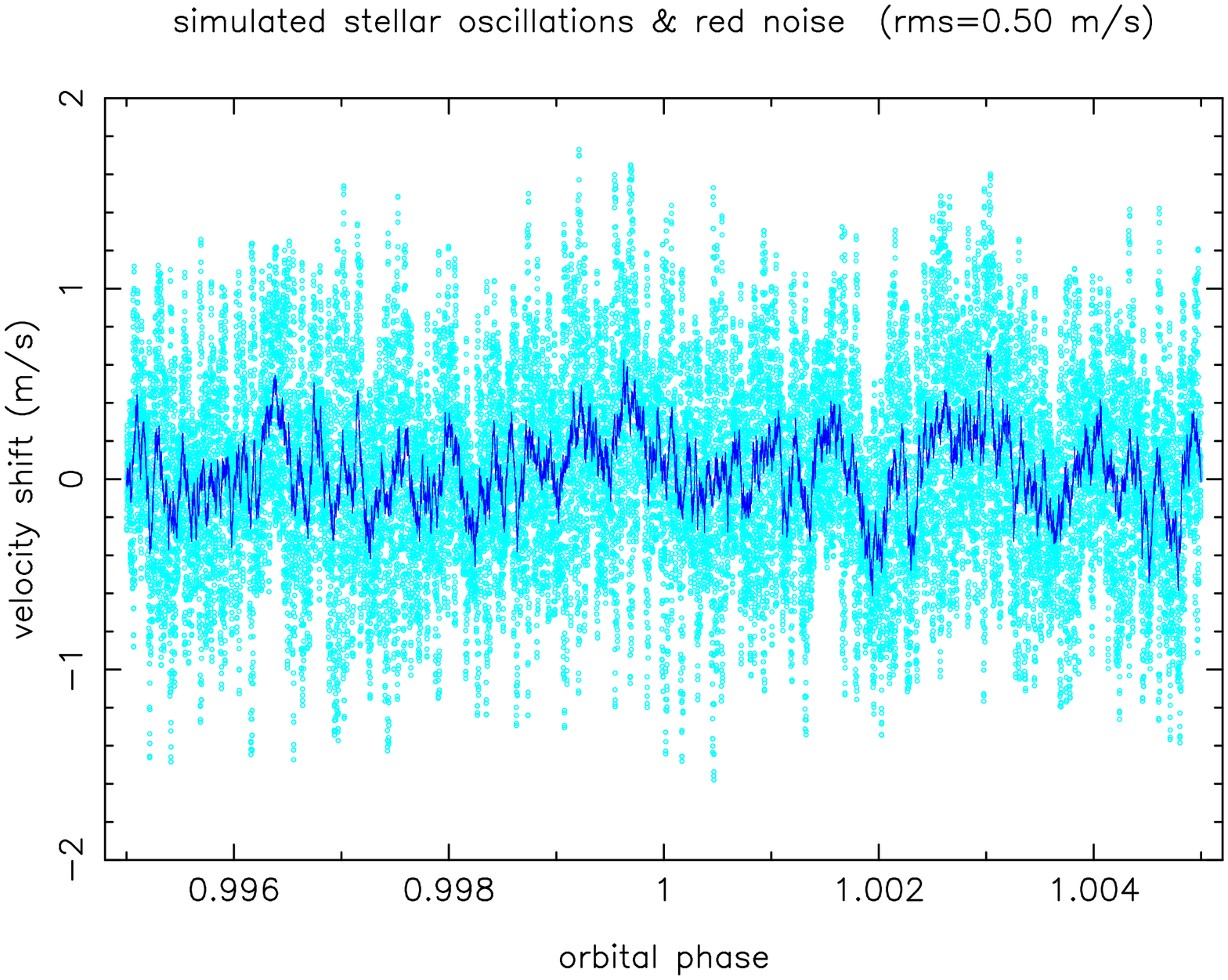}
\end{minipage}
\begin{minipage}[c]{.49\textwidth}
\includegraphics[width=5cm,angle=-90]{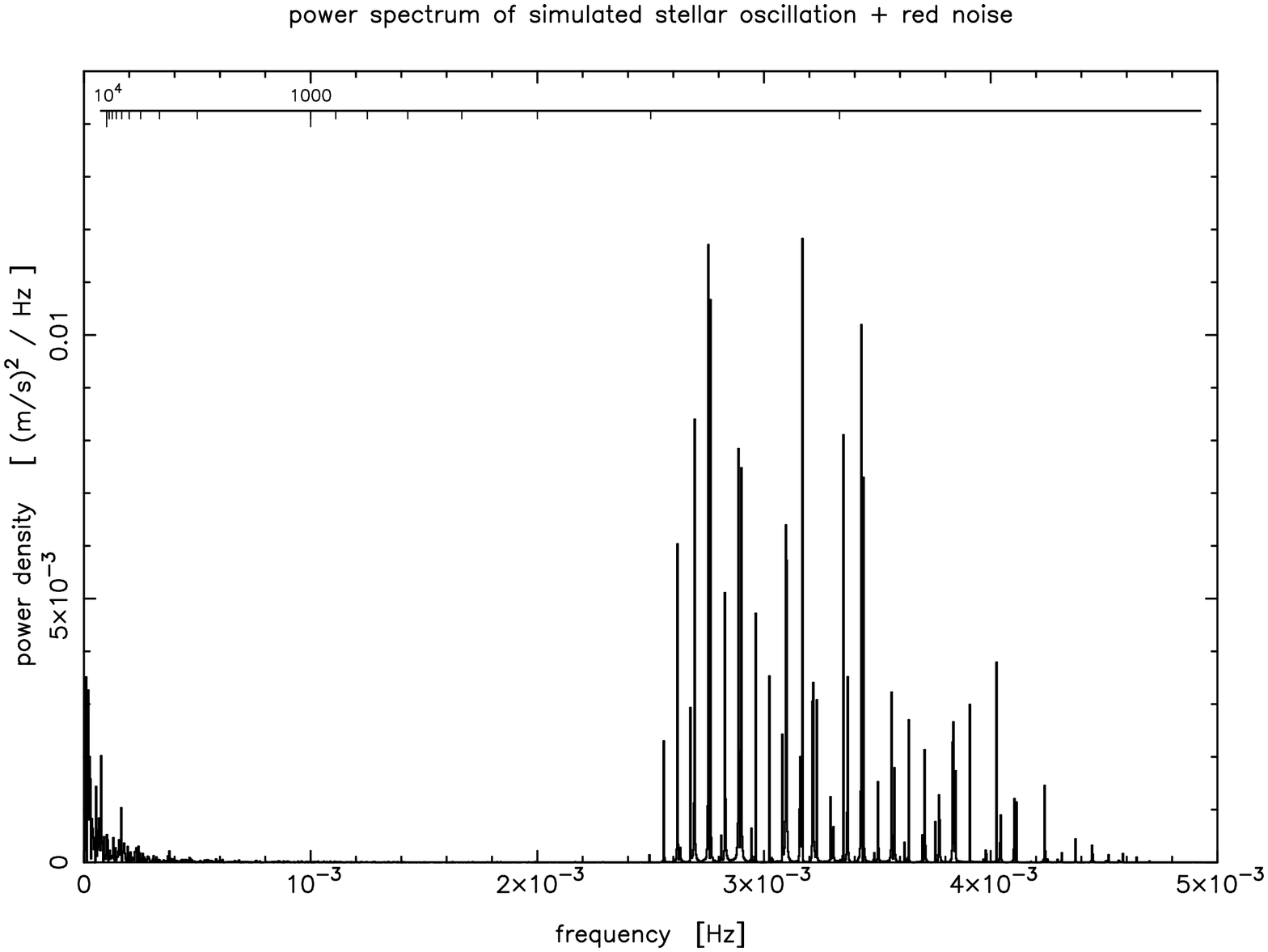}
\end{minipage}
\caption{Simulated velocity shifts and power spectrum at 2 min 
time resolution. No observational noise or Rossiter effect is
present.\label{Fig 1}}
\end{figure}

To compute the Rossiter effect velocities during transit, we use the ELC code 
\citep{wfwOrosz00}
and the semi-analytic prescription given by 
\citet{wfwGimenez2006}. 
We then added the stellar pulsation and burbling noise,
yielding a simulated pure velocity time series with 2 min time resolution.
We then binned to 60~min and added observational noise. We carried out 
several realizations of simulations for each of 9 cases: planet radius 
= 1.0, 1.5 and 2.0 $R_{\earth}$ and $V_{rot} \sin{ i }$ = 3, 4, 5 km/s.
The results are shown in Fig.~2. In this matrix of figures, the lightest curve
shows the ``real'' signal at 2~min resolution, and the dark bins show the 
simulated observations. Notice that when binned to 60~min, the oscillations 
mostly average out and are far less important than the observational and
stellar red noise terms. From these simulations we conclude that
for a 1.0 R$_{\earth}$-size planet, the transit would be undetectable. For
a 1.5 R$_{\earth}$-size planet, the transit might be detectable if $V_{rot} 
\sin{ i } \geq 4$ km/s, and for a 2.0 R$_{\earth}$-size planet the transit 
would easily be detected for $V_{rot} \sin{ i } \geq 3 $ km/s. 
While using the Rossiter effect for discovering an Earth-like planet is 
certainly not competitive with the photometric technique (due to the vast 
multiplexing advantage wide-field photometry has), the Rossiter effect can 
used to provide an important confirmation of potential warm Earth-like 
planets. Since many of the most interesting transits that will be seen by the 
{\it Kepler Mission} will also be the most challenging, an independent 
confirmation of the transit via spectroscopy of the Rossiter effect may prove 
to be very helpful.

\begin{figure}[!ht]
\centering
\includegraphics[width=15.0cm]{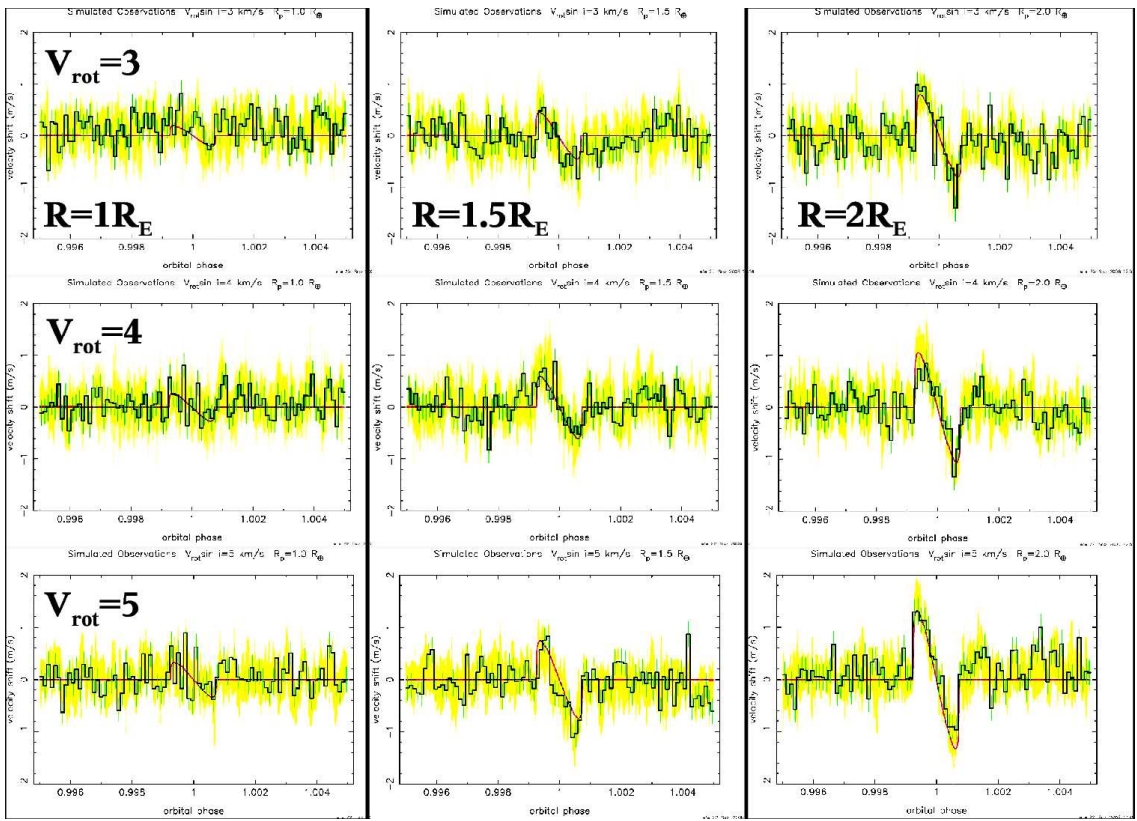}
\caption{Simulations of a transit of an Earth-like planet in a 1 AU orbit
around a Sun-like star.
\label{Fig 2}}
\end{figure}

\acknowledgements 
WFW thanks the University of Texas Astronomy Department for its hospitality
during his 2006 sabbatical visit. Support for this research comes in part from 
a grant from the Research Corporation.


\newpage
\begin{figure}[!ht]
\centering
\includegraphics[height=20.0cm]{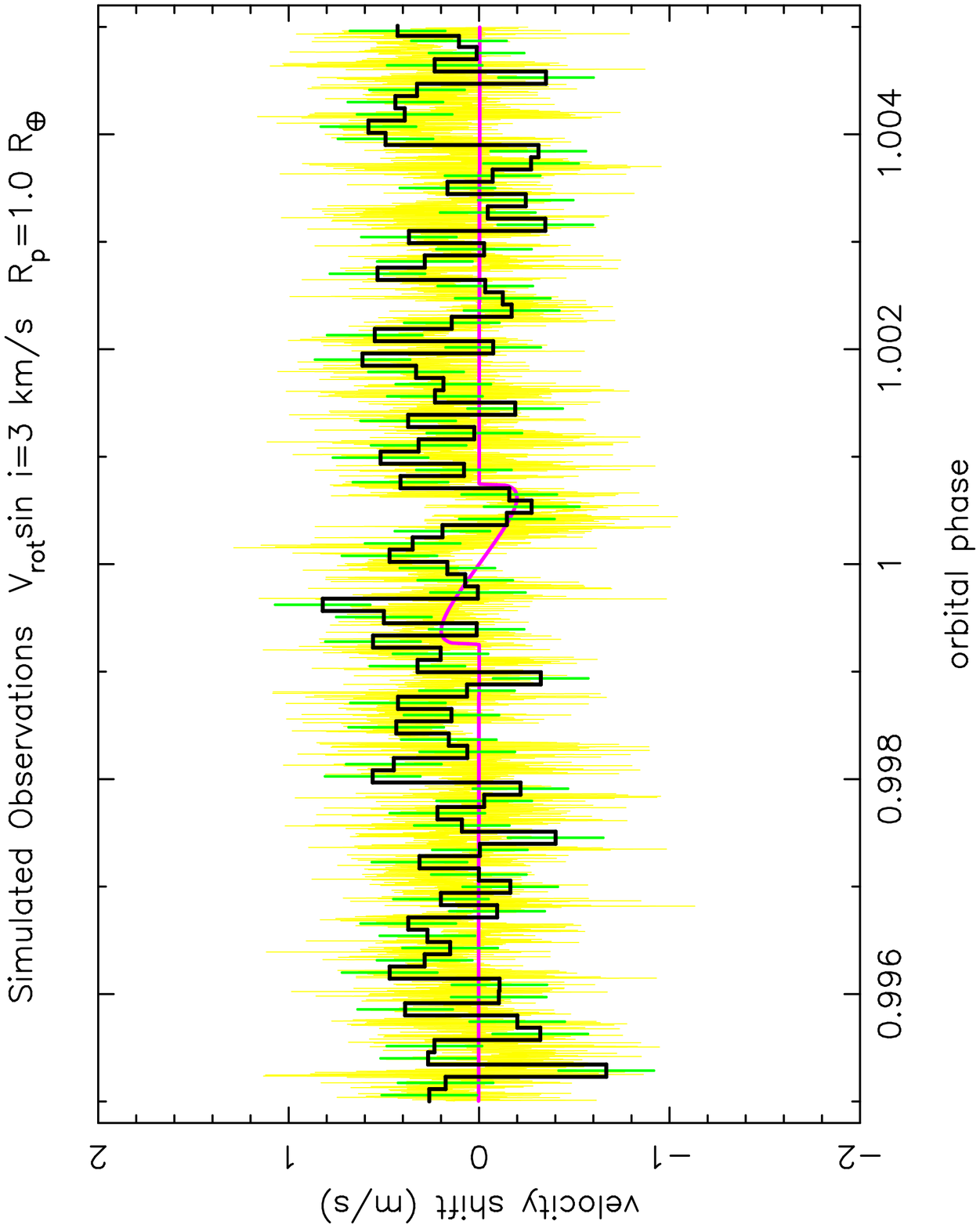}
\end{figure}

\newpage
\begin{figure}[!ht]
\centering
\includegraphics[height=20.0cm]{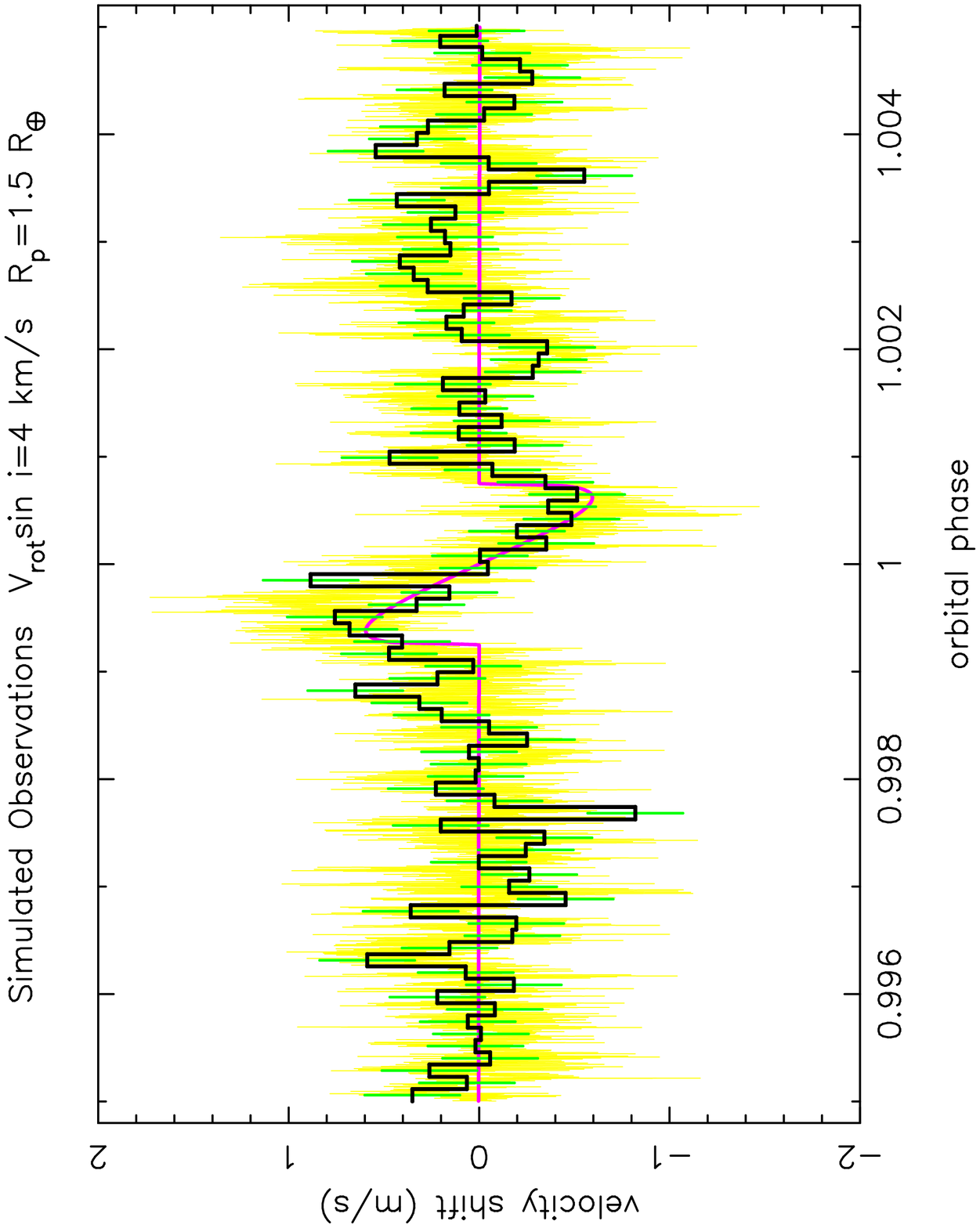}
\end{figure}

\newpage
\begin{figure}[!ht]
\centering
\includegraphics[height=20.0cm]{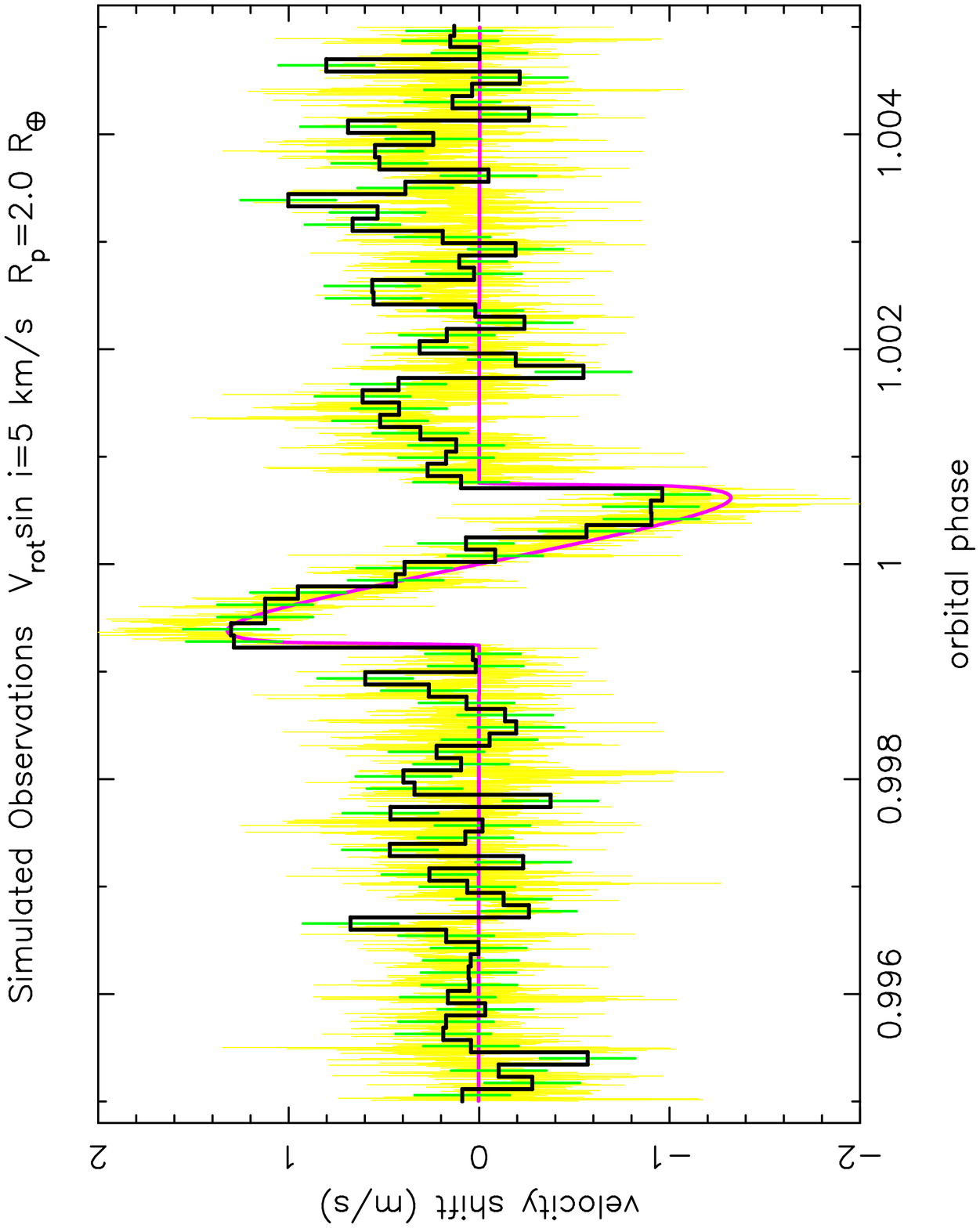}
\end{figure}

\end{document}